%% file: reynolds_main.tex
\documentclass[multphys,vecphys]{svmult}

\usepackage{makeidx}     % allows index generation
\usepackage{graphicx}    % standard LaTeX graphics tool
\usepackage{multicol}    % used for the two-column index
\makeindex             % used for the subject index
\begin{document}

\title*{Magnetohydrodynamic Models for the Structure of Pulsar-Wind Nebulae}
\author{Stephen P. Reynolds}
\institute{North Carolina State University
\texttt{steve\_reynolds@ncsu.edu}}
%\and Name and Address of your Institute \texttt{name@email.address}}
\maketitle

\section{Abstract}

The Crab Nebula is well-described at optical wavelengths and above by
a steady-state magnetohydrodynamic model due to Kennel and Coroniti
(1984).  Can this class of model describe other pulsar-wind nebulae?
I exhibit simple generalizations of KC models, for various values of
$\sigma,$ the ratio of magnetic to particle flux input at the wind
shock.  I calculate the evolution of the electron spectrum and
synchrotron emissivity in the nebula assuming spherical symmetry, a
steady state, and a purely toroidal magnetic field.  Emission profiles
depend on the initial magnetic field $B_0$, the electron spectral
index $s$, and the angle of the toroidal axis with the line of sight,
$\phi$.  I show integrated spectra and radial profiles for various
cases, along with predicted variations of photon index $\Gamma$ with
radius in X-rays.  Most models predict much smaller sizes in X-rays,
and curves of $\Gamma(r)$ which do not resemble observations.  Further
elaborations of the dynamics of PWNe seem necessary.

\section{Introduction}

Pulsar-wind nebulae (PWNe) are bubbles of relativistic particles and
magnetic field inflated by a pulsar, forming center-brightened radio
and X-ray sources of synchrotron radiation \cite{Weiler80},
\cite{RC84}, \cite{KC84a}, \cite{KC84b}.  Most early work on PWNe
assumed they were surrounded by shell supernova remnants (SNRs), seen
or unseen, and interacted with the interior of those remnants.  Before
the passage of a reverse shock, the PWN would expand into unshocked
ejecta in the SNR interior; afterwards, it would be compressed by the
reverse-shock passage and expand much more slowly into the
thermalized, shocked ejecta \cite{RC84}.

The most elaborate early model for the high-energy (optical and above)
emission from PWNe was that of Kennel \& Coroniti \cite{KC84a},
\cite{KC84b} who solved the steady-state relativistic MHD equations in
spherical symmetry with an azimuthal magnetic field to obtain velocity
and density profiles of the outflowing relativistic material.  They
showed that a strong (high compression) MHD shock required a fluid
dominated by particles, not magnetic field, as parameterized by
$\sigma \equiv (B/4 \pi)/n\gamma^2 \beta mc^2$, the ratio of magnetic
(plus electric) energy flux to particle flux just upstream of the wind
shock.  Here $n$ is the comoving density and $\gamma$ the flow Lorentz
factor.  Applying the Rankine-Hugoniot jump conditions at the wind
shock (radius $r_s$) showed that unless $\sigma \ll 1$, the postshock
fluid would remain at high speed, within a factor of 3 of the upstream
(relativistic) speed -- clearly inconsistent with the observed
expansion of the Crab Nebula at $v \sim 2000$ km s$^{-1}$.  Behind the
shock, pure flux-freezing meant that the toroidal magnetic field
strength evolved as $B \propto \rho r$, with $\rho$ the matter
density.  Kennel \& Coroniti \cite{KC84b} were able then to follow the
evolution of electron energies and predict both the integrated
spectrum and the brightness profile of remnants.  They specialized to
the Crab Nebula and showed calculations primarily for that case.  They
were able to describe the optical through hard X-ray spectrum well, as
well as the gross variation of nebular size with X-ray energy
(``size'' being typically the FWHM of emission measured with
relatively crude imaging instruments).  However, the high value of
preshock Lorentz factor of the wind they found ($\gamma \sim 10^6$)
made it difficult to explain the radio emission which requires much
lower energies, and more electrons, than would be produced by
thermalizing this extremely relativistic wind.

Since the advent of the newer generation of X-ray telescopes,
beginning with {\it ASCA} and continuing with the {\it Chandra} X-ray
Observatory and {\it XMM-Newton,} we have seen an explosion in the
number and variety of PWNe in X-rays, and have observed a much wider
range of morphological type.  It is of interest to ask if the class of
steady-state MHD models proposed by KC can be generalized
appropriately to describe other PWNe, and if not, to get an idea of
what kind of extensions might be necessary.  This is the project
described in this paper.

One major problem in modeling several larger, fainter PWNe such as 3C
58 \cite{Torii00} appears in comparing radio and X-ray images.  In
Kennel \& Coroniti's model for the Crab, synchrotron losses on
outflowing particles cause the nebular size to decrease with frequency
at optical and higher frequencies.  In fact the Crab is considerably
smaller at optical than radio frequencies, and continues to diminish
in size roughly as $R \propto \nu^{-0.15}$ up to 40 keV \cite{Ku76}.
However, 3C 58 shows X-ray emission extending to near the edges of the
radio emission, though it is faint there (Slane, private
communication).  These results call into question the MHD picture of
transport of particle energy and magnetic flux throughout the volume
of a PWN.

It is straightforward to show from the MHD equations \cite{KC84a} that
postshock flow solutions have accelerating and decelerating branches,
with the latter beginning at $v_0 = c/3$ and initially decelerating as
$r^{-2}$ (pure hydrodynamic, isobaric flow; magnetic energy is
unimportant since $\sigma$ must be small to fit the outer nebular
boundary conditions).  However, in this case $B \propto \rho r \propto
1/rv \propto r$, so the magnetic energy {\it rises} until it becomes
dynamically important, slowing the deceleration and causing the
velocity to approach a constant value, with $B \rightarrow 1/r$.  The
full hydrodynamic solution can be well approximated by a simple
analytic form: defining $u \equiv v/v_0,$
\begin{equation}
u(r) = {1 \over r^2} (1 - u_\infty) + u_\infty,
\end{equation}
where the asymptotic velocity is $v_\infty/c = \sigma/(1 + \sigma)$ (Fig.~1).

\begin{figure}
\includegraphics[height=4.5cm]{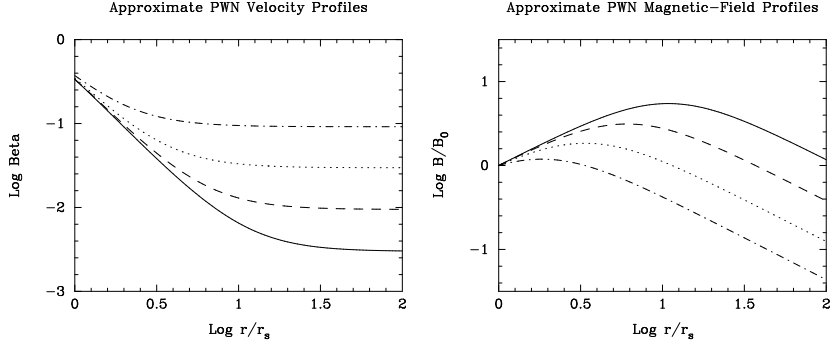}\quad
\includegraphics[height=4.5cm]{reynolds_fig1b.eps}
\caption{Solid lines: $\sigma = 0.001.$ Dashed lines: $\sigma = 0.0032.$
Dotted lines: $\sigma = 0.01.$ Dot-dashed lines: $\sigma = 0.032.$}
\end{figure}

\section{Particle evolution and luminosity calculation}

Electron energies $E$ evolve due both to radiative and adiabatic
losses: $\dot E = (dE/dV) (dV/dt) - a B^2 E^2$, where for synchrotron
losses $a = 1.57 \times 10^{-3}$ cgs.  This equation can be integrated
to give \cite{KC84b}, \cite{R98}
\begin{equation}
E(t) = {E_0 \alpha^{1/3} \over {1 + aE_0 \int B^2 \alpha ^{1/3} dt}}
= {E_0 \alpha^{1/3} \over {1 + E_0 \left( {a B_0^2 r_0 \over v_0} \right)
   \int z^{-8/3} u^{-10/3}dz}}
\end{equation}
where $\alpha \equiv \rho/\rho_0,$ $z \equiv r/r_0,$ and $u \equiv
v/v_0$.  The quantity $E_f \equiv (aB_0^2r_0/v_0)^{-1}$ is a
``fiducial energy'', that an initially infinitely energetic electron
would have after radiating in a field $B_0$ for a time $r_0/v_0$
(basically a parameterization of $B_0$).  From this result, the
evolution of an arbitrary distribution can be calculated:
\begin{equation}
N(E) = N(E_0) {dE_0 \over dE} {dV_0 \over dV} = N(E_0(E)){dE_0 \over dE} \alpha
= N(E_0(E)) \left( {E_0^2 \over E^2} \right) \left( {\rho \over \rho_0} \right)^{4/3}.
\end{equation}

\begin{figure}
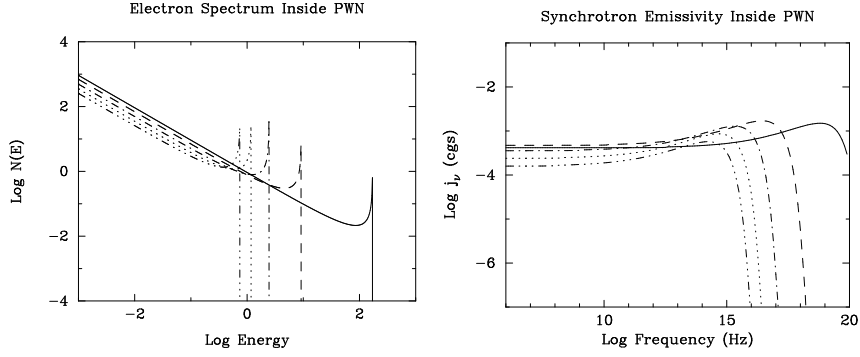

\includegraphics[height=4.6cm]{reynolds_fig2a.eps}\quad
\includegraphics[height=4.5cm]{reynolds_fig2b.eps}
\caption{Flat-spectrum, high-$\sigma$ case: $s = 1.0$, $\sigma = 0.01$,
$E_f = 910$ erg, and $E_{\rm min} = 10^{-5}$ erg.  From left to right, 
lines show the spectrum and emissivity at five radii (outside in):
$r/r_0 = 10, 8, 6, 4, 2$.}
\end{figure}

\begin{figure}
\includegraphics[height=4.5cm]{reynolds_fig3a.eps}\quad
\includegraphics[height=4.5cm]{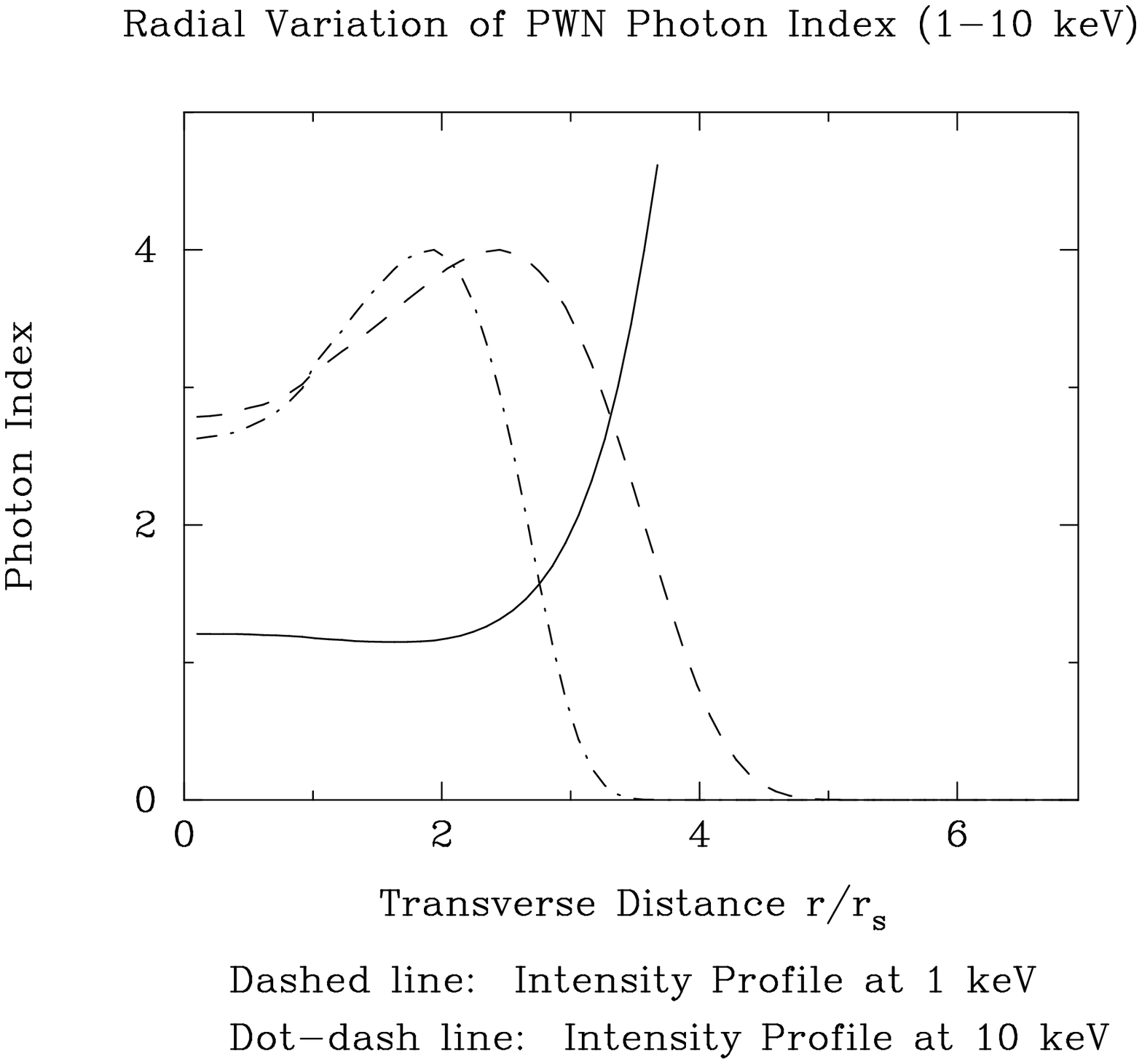}
\caption{Same parameters as above.
Note curvature
of the integrated spectrum due to the energy-loss ``bump''. The solid
line on the right is the 1--10 keV photon index; the profiles are for
aspect angle $\phi = 0^\circ$ ({\bf B} in the plane of the sky).}
\end{figure}

\begin{figure}
\includegraphics[height=4.5cm]{reynolds_fig4a.eps}\quad
\includegraphics[height=4.5cm]{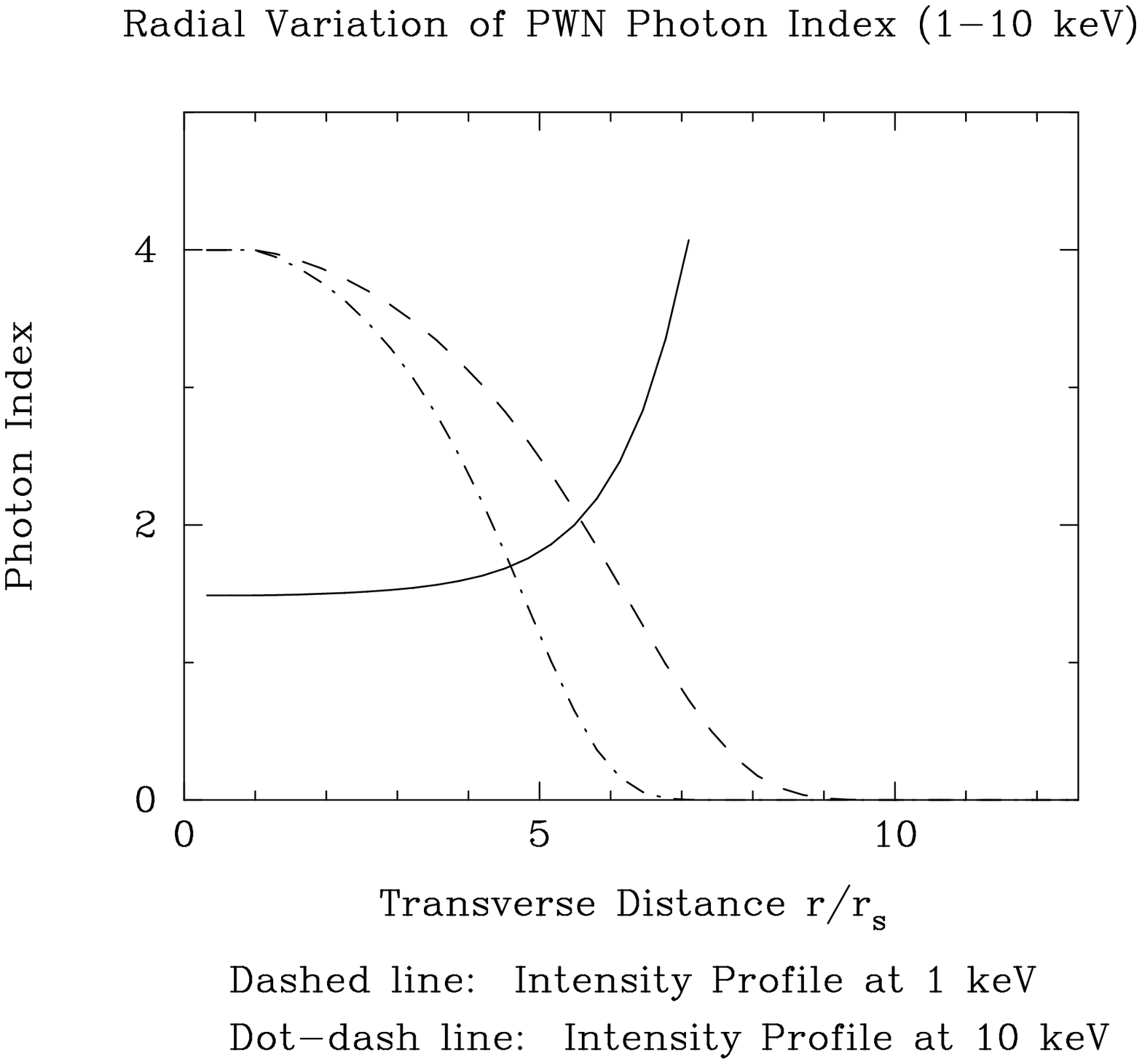}
\caption{Steeper-spectrum, low-$\sigma$ case: $s = 1.5$, $\sigma = 0.001$,
$E_f = 9.1 \times 10^5$ erg, and 
aspect angle $\phi = 90^\circ$.}
\end{figure}

Given the approximate dynamics above, the electron distribution and
synchrotron emissivity can be calculated numerically at each radius.
If magnetic field lines are circles in an equatorial plane, an
elementary rotation can give brightness profiles as a function of
aspect angle $\phi$ between the polar axis and the line of sight.  The
synchrotron emissivity was integrated numerically to generate the
brightness profiles shown above, for an injected flat power-law $N(E)
= KE^{-s}$ with $s < 2$ as deduced from observed PWN radio spectra.
The flat spectra cause a ``bump'' to appear just below the high-energy
cutoff that appears as a result of radiative losses.

\section{Conclusions}

Steady-state MHD models of pulsar-wind nebulae can be well
approximated by a simple velocity law, which then dictates the
evolution of the injected electron distribution and the integrated
spectrum and profile of emission, depending on $\sigma$.  Spectral
properties also depend on {\bf B} and spectral index $s$.  Varying the
aspect angle changes the flux normalization but has no effect on the
integrated spectrum and little on the profile shape at photon energies
for which losses are important.  For very flat injected spectra ($s
\sim 1$), the effects of the energy-loss ``bump'' are significant,
causing concave-up curvature over a broad frequency range.

The models are in significant conflict with the observed radial
dependence of photon index $\Gamma$ in X-rays (e.g., G21.5-0.9
\cite{Slane00}), which appears roughly linear with $r$.  All models
drop in size by at least a factor 2 between radio and X-rays, but some
observed PWNe do not show this.  Unless magnetic-field strengths are
unrealistically low, most X-ray profiles have HWHM radii only 3--6
times the injection radius -- also perhaps at odds with observations.

We conclude that there are significant discrepancies between the
predictions of the simple MHD models and X-ray observations of PWNe.
Nonsteady or nonspherical flows, and/or electron transport by
diffusion as well as convection (e.g., \cite{RJ91}), may be necessary
to explain the observations.

%
%
% BibTeX users please use
% \bibliographystyle{}
% \bibliography{}
%
% Non-BibTeX users please follow the syntax
% the syntax of "referenc.tex" for your own citations
\input{reynolds_ref}

%%%%%%%%%%%%%%%%%%%%%%%%%%%%%%%%%%%%%%%%%%%%%%%%%%%%%%%%%%%%%%%%%%%%%%  }

%%%%%%%%%%%%%%%%%%%%%%%%%%%%%%%%%%%%%%%%%%%%%%%%%%%%%%%%%%%%%%%%%%%%%%

\printindex
\end{document}

%% file: reynolds_ref.tex
%%%%%%%%%%%%%%%%%%%%%%%% referenc.tex %%%%%%%%%%%%%%%%%%%%%%%%%%%%%%
% sample references
% "physics"
%
% Use this file as a template for your own input.
%
%%%%%%%%%%%%%%%%%%%%%%%% Springer-Verlag %%%%%%%%%%%%%%%%%%%%%%%%%%

%
% BibTeX users please use
% \bibliographystyle{}
% \bibliography{}